\documentclass{iopart}
\usepackage{iopams}  
\usepackage{color}
\begin{document}

\title[No-signaling principle and Bell inequality in ${\cal PT}$-symmetric quantum mechanics]{No-signaling principle and Bell inequality in ${\cal PT}$-symmetric quantum mechanics}

\author{George Japaridze, Dipendra Pokhrel, and Xiao-Qian Wang}

\address{Department of Physics, Clark Atlanta University, Atlanta, Georgia 30314, USA}
\vspace{8pt}

\begin{abstract}
${\cal PT}$-symmetric quantum mechanics, the extension of conventional quantum mechanics to the non-Hermitian Hamiltonian invariant under  the combined parity (${\cal P}$) and time reversal (${\cal T}$) symmetry, has been successfully applied to a variety of fields such as solid state physics, mathematical physics,  optics, quantum field theory. Recently,  the extension of ${\cal PT}$-symmetrical theory to entangled quantum systems was challenged  in that ${\cal PT}$ formulation within the conventional Hilbert space violates the no-signaling principle. Here, we revisit the derivation of non-signaling principle in the framework of  ${\cal CPT}$ inner product prescription. Our results preserve the no-signaling principle for a two-qubit system, reaffirm the invariance of the entanglement, and reproduce the Clauser-Horne-Shimony-Holt (CHSH) inequality. We conclude that ${\cal PT}$-symmetric quantum mechanics satisfies the requirements for a fundamental theory and  provides a consistent description of quantum systems.
\end{abstract}
\hspace{25mm}{\it Keywords}: ${\cal PT}$ symmetry, no-signaling, Bell inequality.

\section{Introduction}\label{sec:Introduction}
There are considerable interest in non-Hermitian operators and indefinite Hilbert space structures in quantum mechanics after the discovery that complex Hamiltonians possessing ${\cal PT}$-symmetry (the product of parity and time reversal) can have a real spectrum  \cite{bendb}. The non-Hermitian and ${\cal PT}$-symmetric quantum theory has been successfully used in studies of mathematical physics,  solid state physics, quantum field theory, optics, to mention a few (for a comprehensive literature, see \cite{ review, review2, review3} and references therein).  The diversity of theoretical and experimental investigations call for clarification whether the ${\cal PT}$-symmetric quantum mechanics is a fundamental theory with physically acceptable predictions and conclusions.

The primary requirement that a physically acceptable quantum theory should satisfy, is the reality of the spectrum and the associated conservation of probability. The latter is realized in conventional quantum mechanics by unitarity of the time evolution operator. In this regard, there have been considerable progress in understanding the spectral reality of ${\cal PT}$-symmetric Hamiltonian from establishing the corresponding sufficient conditions and by comparing pseudo-Hermiticity with ${\cal PT}$-symmetry \cite{ review, review2, review3, dorey, most}. The feature of the space of states necessary to define the probability was investigated. The disconcerting problem of the space of states appearing to be an indefinite metric space was subsequently resolved. One solution is based on the observation that self-consistent condition dictates that the space of states is not Hilbert space but  Krein space \cite{gj,znj}. Another solution involves the introduction of the ``charge conjugation'' operator ${\cal C}$ and the resultant ${\cal CPT}$ inner product,  providing a positive defined norm at the outset \cite{b1,b2}. The latter approach logically leads to a unitary time evolution. Thus, the  requirement for a fundamental quantum theory, conservation of the probabilities for Hamiltonians with real spectra is satisfied.

In order to be a fundamental physical theory, ${\cal PT}$-symmetric quantum mechanics should reproduce not only the spectrum reality and unitary but also the general conditions such as no-signaling principle.  The no-signaling principle can be viewed as a realization of classical causality in quantum theory  \cite{nosig, nosign1}. Another essential feature of the fundamental quantum theory  is the violation of Bell inequalities that can be demonstrated as the deviation from the classical probabilistic result for the CHSH thought experiments \cite{cs,cs1,cs2}. In this regard, the no-signaling principle was analyzed in ${\cal PT}$-symmetric quantum mechanics and  it is claimed that the principle is violated \cite{lee}.  The conclusion, based on the calculation in the framework of conventional Hilbert space prescription, implies that ${\cal PT}$-symmetric quantum mechanics cannot be a fundamental theory. 

In this paper, based on the definition of probability via ${\cal CPT}$ inner product, we demonstrate that both no-signaling principle and the result of the CHSH experiment are reproduced by ${\cal PT}$-symmetric extension of quantum mechanics. Therefore, we reaffirm the consistency of ${\mathcal {PT}}$ symmetric quantum mechanics as a  fundamental theory. The paper is organized as follows: in section II we give a brief review of a ${\cal CPT}$ inner product. We demonstrate on the example of a model Hamiltonian that in the space of states endowed with  ${\cal CPT}$ inner product, the orthogonality of states is achieved, and the density matrix is diagonal. The orthogonality is the mathematical realization of the physical requirement,  forbidding any state to have different eigenvalues simultaneously.  In  ${\cal PT}$-symmetric quantum mechanics, Hilbert space formalism fails to satisfy this condition. In section III, using ${\cal CPT}$ inner product prescription, we reexamine the no-signaling principle on the example of two separated observers sharing an entangled state. We illustrate that in both cases, when both observers are equipped with a a non-Hermitian ${\cal PT}$-symmetric Hamiltonians, or when one Hamiltonian is ${\cal PT}$-symmetric and other is Hermitian,  the no-signaling principle is valid. The discrepancy between our results and those from \cite{lee} is attributed to using the different prescriptions for calculating the conditional probabilities. If one uses prescription dictated by ${\cal CPT}$ inner product, the no-signaling principle remains valid. Using ${\cal CPT}$ prescription, we also show that in ${\cal PT}$-symmetric quantum mechanics the measure of entanglement is invariant under unitary transformations, in analogous to the result obtained in conventional quantum mechanics. In section IV, we show that the quantumness, manifested by the CHSH experiment, is reproduced in ${\cal PT}$-symmetric quantum mechanics using  ${\cal CPT}$ inner product. In the last section, we draw our conclusions. Expressions for the eigenfunctions of Hamiltonian for the composite system, density matrix, and  the projection operators necessary for the derivation of our results, are given in Appendices.

\section{ ${\mathcal CPT}$ inner product}
It is well known that the physical requirement that it is impossible to measure simultaneously different eigenvalues \cite{landau,landau1} leads to the conclusion that the space of states of ${\cal PT}$-symmetric quantum mechanics is an indefinite metric space \cite{gj,znj}. The scalar product for two states $\psi_{\alpha}$ and $\psi_{\beta}$, identified with the transition amplitude, satisfying the requirement that it vanishes for the  states with different eigenvalues, $E_{\alpha}\neq E_{\beta}$, is found to be \cite{gj,znj}
\begin{equation}
(\psi_{\alpha}|\psi_{\beta})=({\cal PT}\psi_{\alpha})^{T}\,\psi_{\beta}\,=\,\int\,dx\,\psi_{\alpha}^{\star}(-x)\,\psi_{\beta}(x).
\label{innerr}
\end{equation}
Throughout we consider the real-space independent case; all the results remain intact when wavefunctions depend on $x$. The linear space with indefinite metric described above satisfies the requirement that the transition amplitude between the states with different energies vanishes. Thus, the space equipped with the above scalar product serves as a space of states of ${\cal PT}$-symmetric quantum mechanics. In contrast, the traditional definition of the inner product,  $\langle \psi_{\alpha}|\psi_{\beta}\rangle=(\psi^{T}_{\alpha})^{\star}\,\psi_{\beta}$, violates the orthogonality of states with different eigenvalues. As such, Hilbert space cannot be the  space of states for ${\cal PT}$-symmetric quantum mechanics. Probabilistic interpretation requires  that the amplitude $(\psi_{\alpha}|\psi_{\alpha})$ must be normalizable, which is not  possible in Hilbert space prescription, since the inner product Eq. (\ref{innerr}) is not positively defined. This setback was overcomed by introducing an operator ${\cal C}$, which accounts for the negative sign of the norm defined from Eq. (\ref{innerr}). The inner product in ${\cal PT}$-symmetric quantum mechanics is defined as follows \cite{b1, b2}
\begin{equation}
\label{bbb}
\langle\psi_{\alpha}|\psi_{\beta}\rangle_{{\cal CPT}}=({\cal CPT}\psi_{\alpha})^{T}\,\psi_{\beta}.
\end{equation}
Linear operator ${\cal C}$ that measures the sign of norm (\ref{innerr})  is constructed in terms of the eigenfunctions of Hamiltonian $\psi_{n}$, $H\psi_{n}(x)\,=\,E_{n}\psi_{n}(x)$ \cite{b1},
\begin{equation}
\label{CC1}
{\cal C}=\sum_{n}\psi_{n}\otimes \psi_{n}^{T},
\end{equation}
where ${\cal PT}$ symmetry is assumed not broken.  Here $\otimes$ stands for the tensor product of vectors $\psi_{n}$. The norm associated with Eq. (\ref{CC1}) is positively-defined because it contributes additional factor of -1 when it is applied to the negative vectors $\psi^{-}$ whose ${\cal PT}$ norm is negative: $({\cal PT}\psi^{-})^{T}\,\psi^{-}<0$. In this way, ${\cal C}$ can be considered as the result of measurement of the sign of the inner product $(\psi|\psi)$. The  properties of ${\cal C}$ are similar to those of the charge conjugation operator in quantum field theory \cite{b1}. For the formal properties of operators with ${\cal C}$-symmetries, we refer to \cite{albe} for details. For ${\cal PT}$ and ${\cal CPT}$ frames on a Hilbert space for ${\cal PT}$ symmetric Hamiltonian see \cite{ciao}, \cite{ciao1}.

We now demonstrate how ${\cal CPT}$ inner product prescription works for an example of exactly solvable non-Hermitian ${\cal PT}$-symmetrical Hamiltonian \cite{b1}:
\begin{equation}
\label{Ham}
\fl H_{2\times 2}=s\left(\begin{array}{ccc}
i\sin\alpha&1\\
1&-i\sin\alpha
\end{array}\right), \;{\cal P}=\left(\begin{array}{ccc}
0&1\\
1&0
\end{array}\right),\;{\cal T}\left(\begin{array}{ccc}
x\\
y
\end{array}\right)=\left(\begin{array}{ccc}
x^{\star}\\
y^{\star}
\end{array}\right),
\end{equation}
where $x^{\star}$ is complex conjugate of $x$, $s$ and $\alpha$ are real parameters; $H_{2\times 2}$ is Hermitian when $\alpha=0$. The eigenvalues $E_{\pm}=\pm s\cos\alpha$ are real provided $\alpha<\pi/2$ \cite{b1}.  The corresponding  eigenfunctions of Hamiltonian (\ref{Ham}) are
\begin{equation}
\label{eigg1}
\psi^{+} ={1\over\sqrt{2\cos\alpha}}\left( \begin{array}{ccc}
e^{i\alpha/2}\\
e^{-i\alpha/2}
\end{array}\right), \quad \psi^{-} ={1\over\sqrt{2\cos\alpha}}\left( \begin{array}{ccc}
e^{-i\alpha/2}\\
-e^{i\alpha/2}
\end{array}\right).				 
\end{equation}
According to Eq. (\ref{CC1}), the ${\cal C}$ operator is
\begin{equation}
\label{cop}
{\cal C}\equiv \psi^{+}\otimes {\psi^{+}}^{T}+\psi^{-} \otimes {\psi^{-}}^{T}={1\over\cos\alpha}\left(\begin{array}{ccc}i\sin\alpha & 1\\
				 1 & -i\sin\alpha
				 \end{array}\right).
\end{equation}
It is straightforward to verify that  ${\cal CPT}$ inner product satisfies requirements of the orthogonality and normalization:
\begin{equation}
\label{01}
\langle\psi^{+}|\psi^{-}\rangle_{{\cal CPT}}\,=\,0,\quad \|\psi^{\pm}\|^{2}\,=\,\langle\psi^{\pm}|\psi^{\pm}\rangle_{{\cal CPT}}\,=\,1.
\end{equation}
Note that if one uses the Hilbert space formalism, the orthogonality condition is not satisfied.  Instead of  Eq. (\ref{01}), one now has:
\begin{equation}
\label{added}
(\psi^{+})^{\star}\cdot \psi^-\,=\,\tan\alpha\neq 0.
\end{equation}
Density matrix defined in the space with ${\cal CPT}$ inner product is diagonal:
\begin{equation}
\label{dens}
\fl \quad \quad\quad \rho\,=\,{1\over 2}\biggl(({\cal CPT}\psi^{+})\otimes {\psi^{+}}^{T}\,+\,({\cal CPT}\psi^{-})\otimes {\psi^{-}}^{T}\biggr)={1\over 2}\left( \begin{array}{ccc}
1& 0\\
0 & 1
 \end{array} \right).
\end{equation}
Thus, ${\cal CPT}$ inner product (\ref{innerr}) leads to physically acceptable results: a positively defined metric, orthonormality of eigenfunctions, and diagonal density matrix. 

We introduce an operator ${\cal C}^{\dagger}$
\begin{equation}
\label{Cdagg}
{\cal C}^{\dagger}\equiv {\cal TCT}.
\end{equation}
Since ${\cal T}^{2}=1$ and $[{\cal P},\,{\cal T}]=0$, transformations ${\cal CPT}$ and ${\cal C^{\dagger}P}$ result in a mutually complex conjugated wave functions.  In terms of
\begin{equation}
\label{phi}
\Phi\equiv {\cal C}^{\dagger}{\cal P}\,\psi,
\end{equation}
normalization and orthogonality conditions (\ref{01}) are written in a compact way
\begin{equation}
\label{fff}
\langle\Phi^{\pm}|\psi^{\pm}\rangle  =1,\quad \langle\Phi^{\pm}|\psi^{\mp}\rangle  = 0,
\end{equation}
where the inner product $\langle \Phi|\psi\rangle=\Phi^{\star}\,\psi^{T}$  formally coincides with the prescription of a Hilbert space. The normalized amplitude and the probability of transition from the state described by $\psi_{\alpha}$ to the state described by $\psi_{\beta}$ can be written as:
\begin{equation}
\label{caa}
{\cal A}_{\alpha\beta}={\langle \Phi_{\alpha}|\psi_{\beta}\rangle\over \sqrt{\langle \Phi_{\alpha}|\psi_{\alpha}\rangle}\sqrt{\langle \Phi_{\beta}|\psi_{\beta}\rangle}},\quad p_{\alpha\beta}=\left| {\cal A}_{\alpha\beta}\right|^{2};\quad 0\leq p_{\alpha\beta}\leq 1.
\end{equation}
Eqs. (\ref{fff} - \ref{caa}), along with the fact that in  the space of states endowed by ${\mathcal CPT}$ inner product, the time evolution operator generated by a ${\mathcal PT}$ symmetric Hamiltonian is unitary \cite{b1}, present a self-consistent interpretation of ${\cal PT}$-symmetric quantum mechanics.

\section{No-signaling principle in ${\mathcal PT}$ symmetric quantum mechanics}
Besides satisfying requirements of the spectrum reality and the unitary time evolution, ${\mathcal PT}$ symmetric systems must meet    other basic physical requirements to represent physically acceptable theory. One of such requirements is that they must satisfy the no-signaling principle: given system composed of two subsystems, Alice ($A$) and Bob ($B$), $A$ cannot communicate to $B$, unless $A$ transmits physical information to $B$ \cite{nosig}.

It is well known that quantum mechanics satisfies the no-signaling principle \cite{nosig1}. The thought experiment may be set up as follows: $A$ and $B$ are at different locations and share non separable, entangled quantum state.  $A$ is given a random message from a set of $N$ elements (the outcome of a measurement). $A$ and $B$ perform arbitrary quantum measurements, using their quantum state but without transmission of any physical information. According to the no-signaling principle, $B$ cannot obtain any information on message given to $A$. Regarding the probability distribution, quantum theory satisfies the no-signaling if $B$ cannot infer $A$'s message from outcomes.  The no-signaling condition is formalized as the requirement that conditional probabilities satisfy the following relation \cite{nosig}:
\begin{equation}
\label{nosgdef}
\sum_{a}P(a,\,b|A,\,B)\,=\,P(b|B),
\end{equation}
where $a$ and $b$ are outcomes of measurements $A$ and $B$, performed by Alice and Bob, respectively. $ A$ belongs to a set of possible measurements performed by Alice. If Eq. (\ref{nosgdef}) is not satisfied, then the no-signaling principle is violated.

The validity of the no-signaling principle in ${\cal PT}$-symmetric quantum mechanics, and as a consequence, the status of  ${\cal PT}$-symmetric quantum mechanics as a fundamental theory was recently questioned in \cite{lee,pati}. The thought experiment designed in \cite{lee} consists of two observers Alice and Bob share maximally entangled state $|\psi_{0}\rangle  = (|00\rangle+|11\rangle)/\sqrt{2}$. Alice has a ${\cal PT}$-symmetric non-Hermitian Hamiltonian $H_{2\times 2}$ as described by Eq. (\ref{Ham}), and does not interact with any subsystem on Bob's side. The total Hamiltonian describing the composite system is $H_{4\times 4}=H_{2\times 2}\otimes I_{2\times 2}$, where $I$ is the identity operator, chosen as the Hamiltonian for Bob.

To examine the validity of the no-signaling principle in the set up described above we make  use of similar prescription as for the $2\times 2$ case, normalizing the eigenfunctions based on ${\cal CPT}$ inner product $\langle \Phi_{j} | \psi_{k}\rangle_{{\cal CPT}}$, in a full analogy with Eq. (\ref{phi}), with $|\Phi_{j}\rangle$ being
\begin{equation}
\label{def1}
|\Phi_{j}\rangle\,\equiv\,{\cal C}^{\dag}_{4\times 4}{\cal P}_{4\times 4}|\psi_{j}\rangle,
\end{equation}
where $j=1,\,2,\,3,\,4$, and the eigenfunctions of the total Hamiltonian $H_{4\times 4}$ are given in Appendix. In Eq. (\ref{def1}), operators ${\cal C}^{\dag}_{4\times 4}$ and ${\cal P}_{4\times 4}$ are defined as follows:
\begin{equation}
\label{stars}
{\cal C}^{\dag}_{4\times 4}={\cal C}^{\dag}_{2\times 2}\otimes I_{2\times 2}, \; {\cal P}_{4\times 4}={\cal P}_{2\times 2}\otimes I_{2\times 2},\; {\cal P}_{2\times 2}=\left(\begin{array}{ccc}0 & 1\\
				 1 & 0
				 \end{array}\right),
\end{equation}
and $\psi_{j}$ are eigenfunctions normalized with respect to ${\cal CPT}$ inner product $\langle \Phi_j |\psi_{k}\rangle$.

The eigenvalue equations are
\begin{equation}
\label{eigen}
H_{4\times 4}\,\psi_{1,2}= s\cos\alpha\,\psi_{1,2},\; H_{4\times 4}\,\psi_{3,4} = -s\cos\alpha\,\psi_{3,4},
\end{equation}
and ${\cal P}_{4\times 4}{\cal T}\psi_{j}=\psi_{j}$.

The maximally entangled wave function $\psi_{0}$ that Alice and Bob share after the time evolution gives rise to two possible final outcomes \cite{lee}
\begin{equation}
\label{plus}
\fl \quad \quad \quad \psi^{+}_{f}={1\over{\sqrt{2}\cos\alpha}}\left( \begin{array}{ccc}
\sin\alpha\\
-i\\
-i\\
-\sin\alpha
\end{array}\right), \quad \quad \psi^{-}_{f}={1\over{\sqrt{2}\cos\alpha}}\left( \begin{array}{ccc}
-i\\
\sin\alpha\\
-\sin\alpha\\
-i
\end{array}\right),
\end{equation}
where $\psi^{\pm}_{f}\equiv (U(\tau)A_{\pm}\otimes I)\,\psi_{0}$, $U(\tau)$ is the time evolution operator $U(\tau)\,=\,\exp(-iH_{2\times 2}\,\tau)$ \cite{time}, and $A_{\pm}$ are operators $I$ and $\sigma_{x}$ that Alice uses with respect to the information she wants to send via shared entangled wave function \cite{lee}. The evolution time $\tau$ is set to be $\tau=\pi/(E_+-E_-)$. The possible outcomes $a$ and $b$ are $+y$ or $-y$.

Marginal joint probability, calculated in \cite{lee} using a Hilbert space prescription, results into expression depending on a parameter $\alpha$ in Alice's Hamiltonian
\begin{equation}
\label{lee23}
\fl \quad \quad \sum_{a}P(a,\,b|A,\,B)=\sum_{a=\pm y}\langle \psi^{\pm}_{f}|(|a\rangle\langle a |\otimes |b\rangle\langle b |)|\psi^{\pm}_{f}\rangle={(1\mp
 \sin\alpha)^{2}\over 2 (1+\sin^{2}\alpha)}.
\end{equation}
Eq. (\ref{lee23}) indicates that the Bob's outcomes probabilities depend on what measurement Alice performs, $A_{+}$ or $A_{-}$, which in turn implies the violation of the no-signaling principle. 

Note that it is straightforward to verify that $\psi_{j}$'s are not orthogonal with respect to Hilbert space inner product in that $\langle\psi_{j}|\psi_{k}\rangle\neq\delta_{jk}$. By contrast, ${\cal CPT}$ inner product (\ref{fff}) satisfies the orthonormality condition
\begin{equation}
\label{33}
\langle\Phi_{j}|\psi_{k}\rangle=\delta_{jk}.
\end{equation}
The density matrix is defined the same way as in Eq. (\ref{dens}) and is diagonal
\begin{equation}
\label{dens2}
\rho_{4\times 4}\,=\,{1\over 4}\sum^{4}_{j=1}|\Phi_{j}\rangle\langle \psi_{j}|,\quad(\rho_{4\times 4})_{ij}={\delta_{ij}\over 4}.
\end{equation}
The self-consistent way of formulating ${\cal PT}$-symmetric quantum mechanics requires that the transition between the states with different eigenvalues should vanish. The modulus of amplitude transiting into itself should be normalizable to one.  The latter, in turn, defines the space of states as a linear space with ${\cal CPT}$, rather than Hilbert space  inner product. In our case, when the quantum mechanical system contains subsystem whose dynamics is described by a ${\cal PT}$-symmetric Hamiltonian, the entangled wave function of the system belongs to the space with ${\cal CPT}$ inner product Eq. (\ref{innerr}). All the observables have to be calculated in this space.

Thus, for the entangled states, in analogy to Eq. (\ref{def1}), we introduce $|\Phi^{\pm}_{f}\rangle\equiv\,{\cal C^{\dag}}_{4\times 4}{\cal P}_{4\times 4}|\psi^{\pm}_{f}\rangle$. We then have:
\begin{equation}
\label{final}
|\Phi^{+}_{f}\rangle={1\over \sqrt{2}}\left( \begin{array}{ccc}
0\\
-i\\
-i\\
0
\end{array}\right),\quad |\Phi^{-}_{f}\rangle={1\over \sqrt{2}}\left( \begin{array}{ccc}
-i\\
0\\
0\\
-i
\end{array}\right).
\end{equation}
Orthonormality conditions read as:
\begin{equation}
\label{inpr1}
\langle \Phi^{\pm}_{f}|\Phi^{\pm}_{f}\rangle = 1, \quad \langle \Phi^{\pm}_{f}|\Phi^{\mp}_{f}\rangle = 0.
\end{equation}

We replace  Eq. (\ref{lee23}), the expression for the marginal probabilities considered in \cite{lee}, by
\begin{equation}
\label{13}
\langle \Phi^{\pm}_{f}|(|a\rangle\langle a |\otimes |b\rangle\langle b |)|\Phi^{\pm}_{f}\rangle,
\end{equation}
which respects the physical requirements such as positively defined probability, the orthogonality, and vanishing of the transition amplitude between different states. 
It is worth noting that the projectors $(|a\rangle\langle a |\otimes |b\rangle\langle b |)$ are Hermitian operators. To calculate the marginal probability of a composite system, we  substitute $|\psi^{\pm}_{f}\rangle$ by $|\Phi^{\pm}_{f}\rangle$ in Eq. (\ref{lee23})  and insert the identity $I_{4 \times 4}\,=\,\sum_{j=1}^{4}|\eta_{j}\rangle\langle \eta_{j}|$, where $|\eta_{j}\rangle$ are the eigenvectors of operator $|a\rangle\langle a |\otimes |b\rangle\langle b|$ that filter out the results of the possible outcomes. Then, it is straightforward to verify that
\begin{eqnarray}
\fl \quad \quad \sum_{a}P(a,\,b|A,\,B)&=\sum_{j=1}^{4}\langle \Phi^{\pm}_{f}|a\rangle\,\langle a |\otimes |b\rangle\langle b \,|\eta_{j}\rangle\,\langle \eta_{j}|\Phi^{\pm}_{f}\rangle\,=\\
\label{gd}
&\sum_{a=\pm y} \langle \Phi^{\pm}_{f}|a\rangle\,\langle a |\otimes |b\rangle\langle b \,|\eta_{j=\pm y}\rangle\,\langle \eta_{= \pm y}|\Phi^{\pm}_{f}\rangle\,=\,{1\over 2}.
\end{eqnarray}
As such, Eq. (\ref{gd}) does not contain the parameter $\alpha$. Thus, the no-signaling principle holds in ${\cal PT}$-symmetric quantum mechanics. The probabilities of Bob's outcomes do not depend on the measurements Alice performs.

It is important to note that the calculated probability, Eq. (\ref{gd}), acquires $\Phi^{\pm}_{f}$ at both ends, which is simply attributed to the fact that the probability is the modular square of the amplitude. With the use of ${\cal CPT}$ prescription, the no-signaling principle is preserved. A straightforward calculation verifies the following
\begin{equation}
\label{gd2}
\sum_{a} \langle \Phi^{\pm}_{f}|a\rangle\,\langle a |\otimes |b\rangle\langle b|\Phi^{\pm}_{f}\rangle\,=\,\sum_{b} \langle \Phi^{\pm}_{f}|a\rangle\,\langle a |\otimes |b\rangle\langle b|\Phi^{\pm}_{f}\rangle\,=\,{1\over 2}.
\end{equation}
The same procedure can be applied to the case when Bob also possesses ${\cal PT}$-symmetric Hamiltonian. The Hamiltonian described the corresponding system
\begin{equation}
\label{ham1}
\fl\quad H=H_A(\alpha_A)\otimes H_B(\alpha_B),\;\; H_{A,\,B}(\alpha_{A,\,B})=s_{A,\,B}\left(\begin{array}{ccc}i\sin\alpha_{A,\,B} & 1\\
				 1 & -i\sin\alpha_{A,\,B}
				 \end{array}\right).
\end{equation}
Using eigenfunctions of Hamiltonian (\ref{ham1}) (see Appendix B) it is straightforward to show that the same result (\ref{gd2}) holds, demonstrating that the no-signaling is respected in this case as well. Therefore, no signaling is valid when either one or both Hamiltonians of a system comprised of two subsystems are non Hermitian and ${\cal PT}$-symmetric.

We are now in a position to address an important issue concerning the conservation of entanglement under unitary transformations. It was   stated \cite{pati} that in a two-component system, the measure of entanglement is changed due to the non-Hermiticity of the Hamiltonian (\ref{Ham}). The conclusion poses yet another serious challenge to a ${\cal PT}$-symmetric quantum mechanics since the measure of entanglement, among other general requirements, should be invariant under the local  unitary transformations \cite{nosig,plan}. The measurement of entanglement $E$ is given by \cite{nosig,plan}
\begin{equation}
\label{e}
E(\psi)=-tr_A(\rho_A\,\log\rho_A)=-tr_B(\rho_B\,\log\rho_B),
\end{equation}
where the reduced density matrices are defined as a partial traces: $\rho_A=tr_B(|\psi\rangle\langle \psi|)$ and $\rho_B=tr_A(|\psi\rangle\langle \psi|)$. In \cite{pati} prescription (\ref{e}) is applied to a bipartite system of Alice and Bob described above.  According to \cite{pati}, starting from a  maximally entangled state $\psi_{0}  = (|00\rangle+|11\rangle)/\sqrt{2}$ and using Hilbert space definition for the reduced density matrix, after the time $\tau$, the density operator for Bob's subsystem is
\begin{equation}
\label{bobro}
\rho_B\,=\,{1\over 2}\left(\begin{array}{ccc}1+\sin\alpha\cos\alpha & i\sin\alpha\\
				 -i\sin\alpha & 1-\sin\alpha\cos\alpha
				 \end{array}\right).
\end{equation}
Entanglement measure $E$ is given by
\begin{equation}
\label{E}
E=-\sum_{j}\lambda_{j}\log\lambda_{j},
\end{equation}
where $\lambda_{j}$ are the eigenvalues of the density operator. The eigenvalues of (\ref{bobro}) are $\lambda_{\pm}=(1\pm\sqrt{1-\cos^{4}\alpha})/2$, so $-\sum_{j}\lambda_{j}\log\lambda_{j}$ depends on a non-Hermiticity parameter $\alpha$. After time $\tau$ entanglement measure is no longer unity, which implies that the maximally entangled state changes to a non-maximally entangled one.  The violation of entanglement invariance, if true, would  deprive ${\cal PT}$-symmetric quantum mechanics as a fundamental theory.

Note that the reduced density operator appearing in the expression for the measure of entanglement (\ref{e}) is defined in \cite{pati} as $\rho_B=tr_A(|\psi\rangle\langle \psi|)$, calculated in the framework of the standard quantum theory.  As stated above, the self-consistency of  ${\cal PT}$-symmetric quantum mechanics is achieved when the  space of states is a  linear space equipped with ${\cal CPT}$ inner product in which the density matrix is defined by the prescription (\ref{dens}). ${\cal CPT}$ prescription leads to a physically acceptable result that the no-signaling principle is preserved. 
Therefore, as before, instead of the prescription $|\psi\rangle\langle \psi|$, we use  $|\psi\rangle\langle \Phi|$, where $\Phi\equiv {\cal CPT}\psi$. Note that when $\psi$ is a pure state, the prescription $\rho=|\psi\rangle\langle \psi|$ does not lead to $\rho^2=\rho$ because of the non-orthogonality of $\langle\psi_i|\psi_j\rangle$, while prescription $\rho=|\psi\rangle\langle \Phi|$ results in $\rho^2=\rho$.

We calculate the density matrix using ${\cal CPT}$  inner product and obtain
\begin{equation}
\label{dens1}
\fl \rho={1\over 2}\biggl(|\psi^+\rangle\langle \Phi^+|+|\psi^-\rangle\langle \Phi^-|\biggr)={1\over 4}\left( \begin{array}{cccc}
1 & {2i\sin\alpha\over \cos^{2}\alpha} & 0 & {1+\sin^{2}\alpha\over \cos^{2}\alpha}\\
{2i\sin\alpha\over \cos^{2}\alpha} & 1& {1+\sin^{2}\alpha\over \cos^{2}\alpha} & 0\\
0& {1+\sin^{2}\alpha\over \cos^{2}\alpha} & 1& {2i\sin\alpha\over \cos^{2}\alpha}\\
{1+\sin^{2}\alpha\over \cos^{2}\alpha} & 0 & {2i\sin\alpha\over \cos^{2}\alpha}& 1	
 \end{array} \right)			
 \end{equation}
The reduced density operator for Bob, given by the partial trace of (\ref{dens1}) is
\begin{equation}
\label{dens2}
 \rho_B\,=\,{1\over 2}\left(\begin{array}{ccc}1 & 0\\
				 0 & 1
				 \end{array}\right).
\end{equation}
In distinct of eigenvalues of density matrix (\ref{bobro}), latter calculated using Hilbert space prescription, the eigenvalues of  matrix (\ref{dens2}) are $\lambda_+=\lambda_-=1/2$, and for the entanglement measure we obtain
\begin{equation}
\label{enta}
E=\,-\lambda_+\log \lambda_+\,-\,\lambda_-\log \lambda_-\,=\,1.
\end{equation}
As a result, the entanglement measure remains invariant with regard to unitary transformations in ${\cal PT}$-symmetric quantum mechanics, contrary to claims in \cite{pati}, \cite{prent}.

We demonstrated that both the no-signaling principle and entanglement measure invariance are respected in ${\cal PT}$-symmetric quantum mechanics when the space of states is the space endowed by the ${\cal CPT}$ inner product (\ref{bbb}).

\section{THE CHSH game in ${\cal PT}$-symmetric quantum mechanics}

The CHSH game, introduced by Clauser, Horne, Shimony and Holt \cite{cs}, is a Gedanken experiment devised to manifest the quantumness as the deviation from the classical physics.  The CHSH game is used as a prototype of the Bell test \cite{cs, cs1, cbell}. The experiment is set up as a game where two players, Alice $A$ and Bob $B$, agree on a strategy before  they are separated.  Each of them receives as an  input of a random bit $A,\,B \in (0,1)$ and in response they output a bit each $a,\,b \in (0,1)$. Players win the game when the parity of $a+b$  is equal  to the product of $A$ and $B$, i.e. $a$ and $b$ have to be different from each other if $A=B=1$ and equal to each other for the other three possible combinations of $A$ and $B$.

Evidently, if the players follow a classical strategy, only three of the four possible pairs of inputs  $ (0,0), (0,1), (1,0), (1,1) $ can be satisfied simultaneously. In other words, players ought to agree beforehand that they both output the same bit independently from their input. As such, the maximum winning probability within the classical game is $3/4$, so
\begin{equation}
\label{bell0}
P_{\mathrm{classical}}\,\leq\,0.75,
\end{equation}
which is Bell inequality for the CHSH game.

In the  quantum version of the game, Alice and Bob share an entangled state. Two qubits are separated and $A$ and $B$ measure their qubits depending on the inputs and then output the result of their measurements. Quantum mechanics states that when Alice and Bob start from the maximally entangled state
\begin{equation}
|\psi_{0}\rangle={1\over \sqrt{2}}( |00\rangle+|11\rangle),
\end{equation}
 the maximum winning probability in the quantum game is larger than $3/4$ by agreeing upon a strategy different from the classical one described above,
\begin{equation}
\label{bell1}
P_{\mathrm{quant}}\,\leq\,\cos^2\biggl({\pi\over 8}\biggr)\approx 0.85,
\end{equation}
i.e. the Bell inequality (\ref{bell0}) is indeed violated \cite{cs,cbell}.

In the following, we examine the quantum CHSH game in ${\cal PT}$-symmetric quantum mechanics. The goal is to check whether the Bell inequality is violated in ${\cal PT}$-symmetric quantum mechanics. We use the same procedure as in Hermitian quantum mechanics. Depending on their inputs, Alice and Bob measure their qubits in basis rotated by $0,\,\pi/4,\,\pi/8,\,-\pi/8$, respectively. This rotation is described as follows: when $A=0$, Alice measures in the standard basic $|0\rangle$ and $|1\rangle$ and when $A=1$, Alice measures in the $\pi/4$ rotated basis. Meanwhile Bob measures in the basis rotated by $\pi/8$ when $B=0$ and measures in the basis rotated by $-\pi/8$ when $B=1$. The relative angle between the basis vectors of Alice and Bob is $\pi/8$, $\pi/8$, $\pi/8$, and  $3\pi/8$ for ($A$, $B$)=($0, 0$), ($0, 1$), ($1, 0$ and ($1, 1$), respectively.

The wave function after the time evolution is
\begin{equation}
|\psi^{AB}_{f}\rangle=(U_{\tau}(A)R_{A} \otimes U_{\tau}(B)R_{B})|\psi_0\rangle,
\end{equation}
where $U_{\tau}$  is the time evolution operator, $R_{A}$ and $R_{B}$ are rotating  operators along the basis vectors by angles $\theta_{A}$ and $\theta_{B}$, respectively. 

It is straightforward to verify that without using ${\cal CPT}$ inner product, i.e. if one calculates the marginal probability using $P(ab|AB)=\langle\psi^{AB}_{f}|(|a\rangle\langle a |\otimes |b\rangle\langle b |)| \psi^{AB}_{f}\rangle$, the inequality (\ref{bell1}) is not reproduced.  This conclusion  is similar to the one when the no-signaling principle and  entanglement measure were calculated for ${\cal PT}$-symmetric quantum mechanics using Hilbert space inner product.

For calculation of probabilities, we use prescription (\ref{def1}) which for the time evolved entangled state results in:
\begin{equation}
\label{hohoho}
|\Phi^{AB}_{f}\rangle\equiv {\cal C^{\dagger}P}|\psi^{AB}_{f}\rangle={1\over{\sqrt{2}}}\left( \begin{array}{ccc}
-\cos(\theta_{A}-\theta_{B})\\
\sin(\theta_{A}-\theta_{B})\\
-\sin(\theta_{A}-\theta_{B})\\
\cos(\theta_{A}-\theta_{B})
\end{array}\right).
\end{equation}
Projectors for the measurement of four possible combinations of output bits $a$ and $b$ are
$|a\rangle\langle a|\otimes |b\rangle \langle b|$,     $a\in(0,1),   b\in(0,1)$. Marginal probability calculated using ${\cal CPT}$ inner product is given by
\begin{equation}
P(ab|AB)=\langle\Phi^{AB}_{f}|(|a\rangle\langle a |\otimes |b\rangle\langle b |)| \Phi^{AB}_{f}\rangle,
\end{equation}
and a straightforward calculation leads to
\begin{equation}
\label{ddd}
P(ab|AB)={1\over{2}}\biggl[\cos^2(\theta_{A}-\theta_{B})\delta_{ab}+\sin^2(\theta_{A}-\theta_{B})(1-\delta_{ab})\biggr].
\end{equation}
The winning probability is calculated by summing up marginal probabilities for all possible outputs and inputs.  Alice and Bob agree upon a quantum strategy that the difference between the angles is the same for all the combinations but when $A=B=1$. In other words, we have $\theta_{A}-\theta_{B}=\zeta$ for all cases but for $A=B=1$,
$\theta_{A}-\theta_{B}=3\zeta$ \cite{cs}. Then, the winning probability is
\begin{equation}
\label{winn}
P_{{\mathrm {quant}}}(\zeta)={1\over{4}}[3\cos^2\zeta\,+\,\sin^{2}(3\zeta)].
\end{equation}
After optimization, we obtain  that the winning probability has maximum at $\zeta=\pi/8$ and the maximum is $P^{{\mathrm {max}}}_{{\mathrm {quant}}}=P_{{\mathrm {quant}}}(\pi/8)=\cos^{2}(\pi/8)$. This is in conformity with the CHSH result (\ref{bell1}).

Therefore, in ${\cal PT}$-symmetric quantum mechanics, the probability of winning in quantum CHSH experiment and thus the violation of Bell inequality are reproduced when for the space of states, the linear space equipped with ${\cal CPT}$ inner product is used. It is worth noting that from the Eq. (\ref{ddd}), it follows that $\sum_{a}P(ab|AB)\,=\,1/2$, i.e. the condition for the no-signaling principle is satisfied.
\section{Discussion}
In this paper, we systematically evaluate the entanglement in ${\cal PT}$-symmetric quantum mechanics. The calculations are performed using ${\cal CPT}$ inner product prescription, which is important to rectify the fallacy of using Hilbert space inner product in ${\cal PT}$-symmetric quantum mechanics. Specifically, we demonstrate that the no-signaling principle is valid, the measure of the entanglement is invariant under the unitary transformation, and Bell inequality, presented in the framework of the CHSH experiment, is violated in conformity with results of the conventional quantum mechanics. Therefore, ${\cal PT}$-symmetric quantum mechanics can be a fundamental quantum theory, and offers a sensible description of nature.

At this juncture, it is worth remarking that quantum entanglement plays a pivotal role in the discussion of the no-signaling principle. It is important to clarify the local and non-local correlations associated with quantum entanglement. If we use a separable, untangled initial state such as $|00\rangle$ (a direct product of two bits), the system is referred to as local. If we use an initial entangled state such as $(|00\rangle+|11\rangle)$ the system is non-local and has a violation of the Bell inequality. We showed that in non-local ${\cal PT}$-symmetric systems the no-signaling principle is satisfied only when the physically accepted ${\cal CPT}$ inner product is used. If the initial state is untangled, separable state, the no-signaling principle is satisfied using either Hilbert-space or ${\cal CPT}$-inner product. In other words, in the framework of Hilbert metric, whether the no-signaling principle is violated depends on the existence of quantum entanglement. Using the ${\cal CPT}$ inner product removes this ambiguity: the no-signaling principle is satisfied  in the cases of both local and non-local correlations, as well as for untangled or entangled states.

Our findings pave the way to apply ${\cal PT}$-symmetric quantum correlations to a variety of quantum entangled systems such as quantum information theory, quantum computing, and materials theory. The extension of ${\cal PT}$-symmetric quantum mechanics to entangled systems allows for broad applications. After our work was completed  we learned about recently published papers \cite{brody2}, \cite{znojill} that point to the inappropriate use of the Hilbert space metric in \cite {lee}.  Here we demonstrate explicitly how to employ the ${\cal CPT}$ scheme to quantum probabilities.

\section{ACKNOWLEDGEMENTS}
This research was supported by National Science Foundation (Grant Nos. DMR-0934142 and HRD-1137751).

\appendix
\section{Eigenfunctions, density matrix and the projectors for the composite system $H_{2\times 2}(\alpha)\otimes I_{2\times 2}$.}
The normalized eigenfunctions of the total Hamiltonian $H_{4\times 4}$ are
\begin{equation}
\label{f4}
\psi_{1}={1\over2\sqrt{\cos\alpha} }\left( \begin{array}{ccc}
e^{i\alpha/2}\\
e^{i\alpha/2}\\
e^{-i\alpha/2}\\
e^{-i\alpha/2}
\end{array}\right),\quad \psi_{2}={1\over2\sqrt{\cos\alpha} }\left( \begin{array}{ccc}
e^{i\alpha/2}\\
-e^{i\alpha/2}\\
e^{-i\alpha/2}\\
-e^{-i\alpha/2}
\end{array}\right),
\end{equation}
and
\begin{equation}
\label{f41}
\psi_{3}={1\over2\sqrt{\cos\alpha} }\left( \begin{array}{ccc}
ie^{-i\alpha/2}\\
ie^{-i\alpha/2}\\
-ie^{i\alpha/2}\\
-ie^{i\alpha/2}
\end{array}\right),\quad \psi_{4}={1\over2\sqrt{\cos\alpha} }\left( \begin{array}{ccc}
ie^{-i\alpha/2}\\
-ie^{-i\alpha/2}\\
-ie^{i\alpha/2}\\
ie^{i\alpha/2}
\end{array}\right),
\end{equation}
We follow notations of  \cite{lee}.  For a quantum projectors $|a\rangle\langle a |\otimes |b\rangle\langle b |$ two cases are considered. The first case is the projector $|+y\rangle\langle +y |\otimes |+y\rangle\langle + y|$:
\begin{equation}
\label{proj1}
|+y\rangle\langle +y |\otimes |+y\rangle\langle + y|={1\over 4}\left( \begin{array}{cccc}
1& -i & -i & -1\\
i& 1& 1 & -i\\
i& 1 & 1 & -i\\
-1& i & i & 1
 \end{array} \right),
\end{equation}
and the second case is the projector $|-y\rangle\langle -y |\otimes |+y\rangle\langle +y |$:
\begin{equation}
\label{proj2}
|-y\rangle\langle -y |\otimes |+y\rangle\langle +y |={1\over 4}\left( \begin{array}{cccc}
1& -i & i & 1\\
i& 1& -1 & i\\
-i& -1 & 1 & -i\\
1& -i & i & 1
 \end{array} \right).
\end{equation}
The eigenvalues of these matrices are $1,\,0,\,0,\,0$, respectively. The four eigenvectors $\eta_{j}$ are
\begin{equation}
\label{22}
\fl |\eta_{1}\rangle={ \frac{1}{2} \ }\left( \begin{array}{ccc}
1\\
-i\\
i\\
1
\end{array}\right),\;|\eta_{2}\rangle={ \frac{1}{2} \ }\left( \begin{array}{ccc}
1\\
i\\
i\\
-1
\end{array}\right),\;|\eta_{3}\rangle={ \frac{1}{2} \ }\left( \begin{array}{ccc}
1\\
i\\
-i\\
1
\end{array}\right),\;|\eta_{4}\rangle={\frac{1}{2} \ }\left( \begin{array}{ccc}
-1\\
i\\
i\\
1
\end{array}\right)
\end{equation}
and the eigenvalue equations are:
\begin{equation}
\label{eigeneta}
(|+y\rangle\langle +y |\otimes |+y\rangle\langle +y |)|\eta_{j}\rangle=\delta_{2j}|\eta_{j}\rangle,\end{equation}
and
\begin{equation}
\label{eigeneta2}
(|-y\rangle\langle -y |\otimes |+y\rangle\langle +y |)|\eta_{j}\rangle=\delta_{3j}|\eta_{j}\rangle.
\end{equation}
The density matrix constructed from these states is diagonal:
\begin{equation}
\label{roeta}
\rho_{\eta}={1\over 4}\sum^{4}_{j=1}|\eta_{j}\rangle\langle \eta_{j}|={1\over 4}\left( \begin{array}{cccc}
1& 0 & 0 & 0\\
0& 1& 0 & 0\\
0& 0 & 1 & 0\\
0& 0 & 0 & 1
 \end{array} \right).
\end{equation}

\section{Eigenfunctions  for the composite system $H_{2\times 2}(\alpha_A)\otimes H_{2\times 2}(\alpha_B).$}
\begin{equation}
\label{psi1}
\psi^{++}_{f}={1\over\sqrt{2}\cos\alpha_A\cos\alpha_B}\left( \begin{array}{ccc}
\sin\alpha_A\sin\alpha_B-1\\
-i\sin\alpha_A+i\sin\alpha_B\\
-i\sin\alpha_B+i\sin\alpha_A\\
-1+\sin\alpha_A\sin\alpha_B
\end{array}\right),
\end{equation}

\begin{equation}
\label{psi2}
\psi^{+-}_{f}={1\over\sqrt{2}\cos\alpha_A\cos\alpha_B}\left( \begin{array}{ccc}
-i\sin\alpha_A-i\sin\alpha_B\\
-1-\sin\alpha_A\sin\alpha_B\\
-1-\sin\alpha_A\sin\alpha_B\\
i\sin\alpha_A+i\sin\alpha_B
\end{array}\right),
\end{equation}

\begin{equation}
\label{psi3}
\psi^{-+}_{f}={1\over\sqrt{2}\cos\alpha_A\cos\alpha_B}\left( \begin{array}{ccc}
-i\sin\alpha_A-i\sin\alpha_B\\
-\sin\alpha_A\sin\alpha_B-1\\
-1-\sin\alpha_A\sin\alpha_B\\
i\sin\alpha_A+i\sin\alpha_B
\end{array}\right),
\end{equation}
\begin{equation}
\label{psi4}
\psi^{--}_{f}={1\over\sqrt{2}\cos\alpha_A\cos\alpha_B}\left( \begin{array}{ccc}
\sin\alpha_A\sin\alpha_B-1\\
-i\sin\alpha_A+i\sin\alpha_B\\
-i\sin\alpha_B+i\sin\alpha_A\\
-1+\sin\alpha_A\sin\alpha_B
\end{array}\right).
\end{equation}
Corresponding ${\cal CPT}$ counterparts $\Phi^{ij}_{f}={\cal C^{\dagger}P}\psi^{ij}_{f}$ are:
\begin{equation}
\label{phis}
\fl \Phi^{++}_{f}={1\over\sqrt{2}}\left( \begin{array}{ccc}
-1\\
\;\;0\\
\;\;0\\
-1
\end{array}\right),\, \Phi^{+-}_{f}={1\over\sqrt{2}}\left( \begin{array}{ccc}
\;\;0\\
-1\\
-1\\
\;\;0
\end{array}\right),\,\Phi^{-+}_{f}={1\over\sqrt{2}}\left( \begin{array}{ccc}
\;\;0\\
-1\\
-1\\
\;\;0
\end{array}\right),\, \Phi^{--}_{f}={1\over\sqrt{2}}\left( \begin{array}{ccc}
-1\\
\;\;0\\
\;\;0\\
-1
\end{array}\right).
\end{equation}


\end{document}